\documentclass[conference]{IEEEtran}

\usepackage{cite}
\usepackage{amsmath,amssymb,amsfonts}\usepackage{booktabs}

\usepackage{pifont}
\usepackage{placeins}
\usepackage{multirow}
\usepackage{graphicx}
\usepackage{booktabs}
\usepackage{array}
\usepackage{multirow}
\usepackage{url}
\usepackage{booktabs}
\usepackage{pifont}
\usepackage{hyperref}
\usepackage{xcolor}
\usepackage{float}
\usepackage{algorithm}
\usepackage{algpseudocode}
\usepackage[table]{xcolor}
\begin{document}

\title{Measurement Study of Post-Quantum Readiness of Internet:  2026}
\author{
\IEEEauthorblockN{Vanishka Mohan Dubey}
\IEEEauthorblockA{
Independent Researcher \\
Jammu, India  \\
vanishkadubey@gmail.com}

\and

\IEEEauthorblockN{Gaurav Varshney
}
\IEEEauthorblockA{
Department of CSE \\
IIT Jammu \\
Jammu, India \\
gaurav.varshney@iitjammu.ac.in}
}
\maketitle

\begin{abstract}
The emergence of quantum computing presents a fundamental challenge to the security of current Internet communication systems. Transport Layer Security (TLS), which forms the backbone of secure web communication, predominantly relies on classical public-key cryptographic algorithms such as RSA and elliptic curve cryptography (ECC), both of which are susceptible to quantum attacks. This paper conducts a large-scale empirical evaluation of post-quantum readiness across 32,011 domains, with a primary focus on real-world TLS deployments across diverse sectors by analyzing negotiated TLS parameters, including protocol versions, cipher suites, key exchange mechanisms, and certificates. The results indicate that while modern protocols like TLS 1.3 and QUIC are gaining adoption, 15.70\% of domains especially in critical sectors such as banking and government still rely on TLS 1.2.

Furthermore, 49.3\% of domains support hybrid post-quantum key exchange mechanisms (e.g., ML-KEM-768 with X25519), whereas 50.7\% continue to use classical key exchange, reflecting partial transition. Notably, 0\% adoption of hybrid post-quantum certificates was observed, leaving the authentication layer vulnerable to quantum-enabled attacks such as certificate forgery. The findings reveal uneven adoption of post-quantum mechanisms across sectors, where technology-driven platforms are advancing more rapidly than legacy-dependent infrastructures. Overall, the study highlights that achieving complete quantum resilience requires a coordinated transition not only in key exchange mechanisms but also in certificate infrastructures. Without such comprehensive migration, Internet communication systems remain vulnerable to long-term threats, including Harvest-Now-Decrypt-Later (HNDL) attacks.
\end{abstract}

\begin{IEEEkeywords}
 Quantum Security, Post-Quantum Cryptography, TLS, Internet Measurement, Hybrid Cryptography, key exchanges,TLS certificates. 
\end{IEEEkeywords}
\section{Introduction}

Transport Layer Security (TLS) is the primary protocol used to secure communication across the modern Internet. It protects sensitive information such as financial transactions, personal data, login credentials, and government records by providing confidentiality, authentication, and integrity~\cite{alnahawi2023pqtls}. The emergence of quantum computing, however, introduces a long-term challenge to the cryptographic foundations of TLS.

Many widely deployed public-key mechanisms, including RSA and Elliptic Curve Cryptography (ECC), derive their security from mathematical problems that are considered computationally infeasible for classical computers~\cite{alnahawi2023pqtls}. RSA security relies on the hardness of integer factorization, while ECC relies on the elliptic curve discrete logarithm problem. These assumptions may no longer hold in the presence of large-scale quantum computers. In particular, Shor’s algorithm~\cite{shor1994,vadisetty2024quantum} can efficiently solve both integer factorization and discrete logarithm problems in polynomial time, enabling a quantum adversary to recover private keys used in TLS handshakes.

In addition, symmetric cryptographic primitives such as AES, used within TLS, may experience reduced security under quantum computation. Grover’s algorithm provides a quadratic speedup for brute-force search, effectively reducing the security strength of symmetric ciphers and hash functions. For example, a symmetric key with 128-bit security provides approximately 64 bits of effective security against a quantum attacker~\cite{vadisetty2024quantum}.

TLS forms the foundation of secure communication on the Internet and is widely deployed across web applications, cloud platforms, and content delivery infrastructures. As a result, any weakness in its cryptographic design or implementation can have large-scale implications for Internet security.

In this context, it becomes essential to evaluate TLS under a quantum threat model. This study aims to identify quantum-vulnerable cryptographic primitives, detect the presence of post-quantum or hybrid algorithms, and analyze the real-world deployment of TLS across different sectors. Furthermore, the study evaluates TLS configurations in web servers and Content Delivery Networks (CDNs), which play a critical role in TLS termination, traffic distribution, and security enforcement.
The analysis also considers the impact of quantum threats on the fundamental principles of information security, namely confidentiality, integrity, and authenticity (CIA triad), within modern Internet communication systems.

This approach enables a focused and comprehensive assessment of TLS-based quantum security readiness across Internet infrastructure, particularly at the web and CDN layers.
\subsection{Need of Post-Quantum cryptography}

Post-quantum cryptography (PQC) refers to cryptographic algorithms designed to protect digital systems against potential attacks from quantum computers. ~\cite{nist_pqc, shor1994, alnahawi2023pqtls} . According to NIST, PQC provides a defense mechanism against future quantum-enabled adversaries by replacing classical public-key cryptographic schemes with quantum-resistant alternatives. Unlike quantum cryptography, PQC can be implemented on existing classical systems without requiring specialized quantum hardware. NIST has initiated a global standardization effort and released initial PQC standards to support secure communication in the post-quantum era.
The transition to PQC is essential as current cryptographic systems are not secure against large-scale quantum attacks, and migration to new standards requires significant time and effort.

\section{Related Work}

Several prior studies have conducted large-scale measurement analysis of Internet security protocols, particularly focusing on TLS deployments.

\newcommand{\cmark}{\ding{51}}
\newcommand{\xmark}{\ding{55}}
\begin{table*}[htbp]
\centering
\caption{Comprehensive Comparison of Existing Studies and Our work}
\label{tab:relatedworkcomparison}

\renewcommand{\arraystretch}{1.3}

\resizebox{\textwidth}{!}{
\begin{tabular}{p{3.2cm}ccccp{5.8cm}cc}
\toprule

\textbf{Study} &
\textbf{Measurement} &
\textbf{Dataset} &
\textbf{Real-World} &
\textbf{Sector-wise} &
\textbf{Parameters } &
\textbf{Cryptographic} &
\textbf{HNDL Attack} \\

&
\textbf{}&
\textbf{Size} &
\textbf{Analysis} &
\textbf{Analysis} &
&
\textbf{Analysis} &
\textbf{Risk Analysis} \\

\midrule

Holz et al.~\cite{holz2016tls}
& \cmark
& Millions
& \cmark
& \xmark
& TLS versions, cipher suites, key exchange mechanisms, certificate chains, cryptographic configurations
& Classical TLS security analysis
& \xmark \\ \\

Durumeric et al.~\cite{durumeric2017https}
& \cmark
& 7.75 Billion Handshakes
& \cmark
& \xmark
& HTTPS deployment, TLS versions, cipher suite configurations, PKI ecosystem, certificate validation
& Classical Internet-scale TLS security
& \xmark \\ \\

Benitez~\cite{benitez2026quantum}
& \xmark
& N/A
& \xmark
& \xmark
& TLS, SSH, IPsec, DNSSEC, certificate algorithms, cryptographic infrastructures, vulnerable asymmetric primitives
& Post-Quantum cryptographic risk assessment
& \cmark \\ \\

Raavi et al.~\cite{raavi2025pqc}
& \xmark
& Experimental Setup
& \xmark
& \xmark
& PQC-enabled TLS, PQC certificates, hybrid PKI architectures, certificate validation mechanisms
& PQC integration and security analysis
& \cmark \\ 

\\

Alnahawi et al.~\cite{alnahawi2023pqtls}
& \xmark
& Experimental Setup
& \xmark
& \xmark
& TLS handshake performance, PQC cipher suites, hybrid key exchange, latency and computational overhead
& Post-Quantum TLS performance and security analysis
& \cmark \\ \\

Holz et al.~\cite{holz2019tls13}
& \cmark
& 275 Million Hosts
& \cmark
& \xmark
& TLS 1.3 deployment, handshake behavior, cipher suite adoption, protocol transition analysis
& Classical TLS deployment security
& \xmark \\ \\

Chung and Vlajic~\cite{chung2022tls}
& \cmark
& 50 Banking Domains
& \cmark
& Banking Sector
& TLS configurations, certificate properties, protocol versions, banking-sector cryptographic deployment
& Classical TLS security evaluation
& \xmark \\
\\ 

Sowa et al.~\cite{sowa2024post}
& \cmark
& 13 TB Metadata, 20.5 Million SSH Connections
& \cmark
& \xmark
& PQC adoption rates, SSH/TLS cipher suites, OpenSSH PQC deployment, hybrid key exchange mechanisms, cryptographic metadata analysis
& Post-Quantum cryptographic deployment and migration analysis
& \cmark \\
\\

\textbf{This Study}
& \cmark
& \textbf{32,011 Active Internet Domains}
& \cmark
& \cmark
& \textbf{TLS versions, cipher suites, key exchange mechanisms, certificate algorithms, cryptographic configurations, PQC readiness, CDN providers, web-server providers, sector-wise adoption trends, security posture evaluation}
& \textbf{Classical + Post-Quantum cryptographic analysis}
& \cmark \\
\\ 
\bottomrule
\end{tabular}
}
\end{table*}
Holz et al.~\cite{holz2016tls} performed one of the earliest comprehensive analyses of TLS usage on the web, examining certificate validity, cipher suite configurations, and protocol versions across millions of domains. Their study highlighted widespread use of outdated cryptographic mechanisms and misconfigurations in TLS deployments. However, the analysis was limited to classical security considerations and did not evaluate the implications of emerging quantum threats.

Durumeric et al.~\cite{durumeric2017https} conducted large-scale Internet-wide TLS scanning  analyzing certificate chains, protocol versions, and cryptographic configurations. However, their work focuses on classical security properties and does not evaluate quantum vulnerabilities or post-quantum readiness. 

Benitez~\cite{benitez2026quantum} presents an engineering inventory of cryptographic dependencies across real-world systems, identifying protocols such as TLS, SSH, IPsec, and DNSSEC that rely on quantum-vulnerable primitives like RSA, ECC, and Diffie–Hellman. The study systematically maps quantum threat vectors, including Harvest-Now-Decrypt-Later (HNDL) and signature forgery, across multiple technology domains and operational environments. However, the work is primarily taxonomy-based and does not include large-scale empirical measurement or real-world Internet scanning of cryptographic deployments.

Raavi et al.~\cite{raavi2025pqc} analyze the security and performance of post-quantum digital signature algorithms (Dilithium, Falcon, and SPHINCS+) and their integration with TLS and PKI, focusing on computational cost and handshake overhead. However, their work is limited to controlled environments and does not consider large-scale Internet measurements or real-world deployment. 

Alnahawi et al.~\cite{alnahawi2023pqtls} provide a systematic study of post-quantum TLS handshake mechanisms, classifying existing approaches into post-quantum authentication, key exchange, and hybrid models, along with performance evaluations of PQC algorithms. The study highlights that classical TLS primitives such as RSA, ECDSA, and Diffie–Hellman are vulnerable to quantum attacks, enabling decryption of past sessions and signature forgery. However, the work is primarily survey- and evaluation-based and does not include large-scale Internet measurement or real-world deployment analysis.

Holz et al.~\cite{holz2019tls13} conducted a large-scale measurement study of TLS~1.3 deployment using active scanning across more than 275 million domains and passive monitoring of real-world traffic. Their work analyzes protocol adoption, usage patterns, and the role of major cloud providers in accelerating TLS~1.3 deployment. However, the study focuses on protocol evolution and deployment trends and does not evaluate quantum security, post-quantum cryptographic mechanisms, or multi-protocol environments.

Sowa et al.~\cite{sowa2024post} presented the first large-scale network instrument for monitoring post-quantum cryptography (PQC) adoption across multiple network protocols including SSH, TLS, RDP, and DNS. Their study analyzed real-world traffic at a national-scale supercomputing center using Zeek-based monitoring and identified early adoption of hybrid PQC mechanisms such as sntrup761x25519-sha512 in OpenSSH. The work also highlighted migration challenges, protocol readiness, and the risks of weak cryptographic configurations under emerging quantum threats. However, the study primarily focuses on scientific and HPC environments and does not provide real world sector-wise large-scale Internet analysis across domains such as banking, government, defence, and commercial web infrastructure etc.

Shahwar et al.~\cite{shahwar2024qc} conducted a systematic literature review on quantum cryptography, analyzing 134 studies to identify key challenges, applications, and advancements in QKD and quantum-secured communication. The study highlights the limitations of classical cryptographic schemes against quantum attacks and emphasizes the need for secure future network protocols. However, the work is survey-based and does not include empirical measurement or large-scale analysis of real-world cryptographic deployments.

Chhetri et al.~\cite{chhetri2025pqc} present a comprehensive survey of post-quantum cryptography, covering major algorithmic families such as lattice-, code-, hash-, multivariate-, and isogeny-based schemes, along with their security assumptions, performance trade-offs, and standardization status. The study also discusses practical aspects including protocol integration (e.g., TLS, DNSSEC), hybrid migration strategies, and deployment challenges across domains such as cloud and IoT. However, the work is survey-oriented and does not perform empirical measurement or large-scale analysis of real-world cryptographic deployments. 

Chung and Vlajic  et al~\cite{chung2022tls} examined the real-world deployment and vulnerabilities of TLS in critical sectors. conducted a comprehensive analysis of TLS security in the world’s 50 largest banking websites using remote scanning techniques. Their findings reveal that less than 50\% of the analyzed banks support TLS 1.3, while 94\% still rely on TLS 1.2, exposing them to multiple known vulnerabilities. Furthermore, a significant portion of these systems were found to support insecure cryptographic configurations such as CBC-based cipher modes, making them vulnerable to attacks including BEAST, SWEET32, and Lucky13. Notably, approximately 78\% of the surveyed banking websites were specifically vulnerable to the Lucky13 attack, highlighting the persistent security risks in real-world TLS deployments.While the study by Chung and Vlajic provides important insights into real-world TLS vulnerabilities in the banking sector, it is limited to classical security analysis and focuses on a relatively small dataset of 50 domains. Their work primarily evaluates protocol weaknesses, legacy cipher usage, and known attacks such as BEAST and Lucky13, without considering emerging quantum threats. While this work provides valuable insights into TLS vulnerabilities and deployment practices in the banking sector, it does not consider emerging quantum threats or the adoption of post-quantum cryptographic mechanisms in modern Internet infrastructure.

Nouioua et al.~\cite{nouioua2021quantum} analyze the impact of quantum computing on information security, but do not perform real-world measurement or deployment analysis.
 
Recent works have explored TLS ecosystem evolution, including the adoption of TLS 1.3 and modern cipher suites. However, these analyses still rely on classical threat models and do not assess vulnerabilities arising from quantum computing, particularly in key exchange and certificate algorithms.

In parallel, research on post-quantum cryptography has focused on algorithm design and experimental deployment. Google and Cloudflare~\cite{googlepqc} demonstrated hybrid key exchange mechanisms combining X25519 with post-quantum algorithms such as Kyber.
While prior work has demonstrated the feasibility of integrating post-quantum cryptography (PQC) into TLS, it has largely been conducted in controlled environments and lacks large-scale empirical analysis of real-world adoption across diverse sectors. Moreover, despite extensive research on TLS security and PQC, existing studies often overlook quantum threat models in practical Internet deployments, particularly in terms of large-scale evaluation of post-quantum and hybrid cryptographic mechanisms and their adoption across different sectors as shown in Table \ref{tab:relatedworkcomparison}. To address this gap, this study presents a comprehensive, sector-wise empirical analysis of Internet security across \textbf{32,011 domains} spanning critical sectors such as banking, government, defence, and cloud platforms. Specifically, the work conducts a large-scale measurement of TLS deployments and examines their cryptographic configurations under a quantum threat model, providing a structured assessment of post-quantum readiness across sectors.

\subsection{Contributions}

This paper presents a comprehensive empirical study of post-quantum cryptographic (PQC) readiness in real-world Internet infrastructure. The key contributions are as follows:

\begin{itemize}

\item \textbf{Large-scale Internet measurement under a quantum threat model:}  
We analyze 32,011 domains across multiple critical sectors, including banking, government, defence, and cloud platforms, to evaluate real-world TLS deployments and their exposure to quantum attacks.

\item \textbf{Multi-layer TLS measurement and validation framework:}  
We design an automated framework integrating Selenium-based browser inspection, OpenSSL probing, and Wireshark-based packet validation to extract accurate negotiated TLS parameters.

\item \textbf{Quantitative evaluation of hybrid post-quantum key exchange adoption:}  
We measure the deployment of hybrid mechanisms (e.g., X25519 + ML-KEM-768) or or post quantum mechanisms.

\item \textbf{Identification of a critical gap in certificate-based authentication:}  
We demonstrate that whether analyzed domains use post-quantum or hybrid certificate signature algorithms,.

\item \textbf{Sector-wise quantum risk analysis:}  
We provide a detailed sector-wise assessment highlighting uneven PQC adoption, where cloud and technology platforms show higher readiness compared to critical infrastructure sectors such as BFSI, government, and defence.

\item \textbf{Analysis of TLS negotiation and fallback behavior:}  
We show that client–server negotiation can lead to fallback from PQC-capable configurations to classical cryptography, resulting in inconsistent security guarantees.

\end{itemize}
\section{Data Collection, Methodology, and Validation Framework}

The data collection process was performed using a combination of automated scripting, browser-based analysis, and packet-level validation to ensure accuracy and reliability of the results.

The experiments were conducted using Multiple tools and techniques were used to extract and validate TLS and protocol-level security parameters.

\begin{table*}[htbp]
\centering
\caption{Example of Sector-wise Distribution of 32,011 Domains}
\label{tab:sector_domains}
\renewcommand{\arraystretch}{1.1}
\setlength{\tabcolsep}{3pt}
\scriptsize
\begin{tabular}{p{4cm} p{4cm} p{9.5cm}}
\toprule
\textbf{Sector} & \textbf{Domains}  & \textbf{Example Domains} \\
\midrule

Banking \& Finance (BFSI) & 183 &
sbi.bank.in, hdfc.bank.in, icici.bank.in, axis.bank.in, kotak.bank.in, unionbankofindia.bank.in, yes.bank.in, rbl.bank.in, idfcfirst.bank.in, pnb.bank.in, jkb.bank.in, citi.com, hsbc.bank.in, airtelpayments.bank.in, paytm.bank.in tbank.ru, sberbank.ru, citizensbank.com, otpbank.ru, halykbank.kz, akbank.com, pochtabank.ru, akbank.com, bankofscotland.co.uk, hellobank.fr, kiwibank.co.nz,sbishinseibank.co.jp,apobank.de, primbank.ru, rabobank.com.au,bankofireland.com, alfabank.ru, southstatebank.com, reisebank.de, bankofcyprus.com, bcs-bank.ru, landbank.com, axosbank.com,privatbankar.hu, bankofamerica.com
etc \\
\\
Government \& Public Services & 921 &
india.gov.in, passportindia.gov.in, pmindia.gov.in, niti.gov.in, upsc.gov.in, drdo.gov.in, mha.gov.in, mea.gov.in, pib.gov.in, gst.gov.in etc \\
\\
Defence \& Military & 328 &
drdo.gov.in, mod.gov.in, joinindianarmy.nic.in, nsg.gov.in, crpf.gov.in etc \\
\\
IT \& Software / Technology Platforms & 241 &
tcs.com, infosys.com, wipro.com, microsoft.com, oracle.com, ibm.com, zoho.com, browserstack.com etc \\
\\
Cloud / Hosting / Security & 468 &
aws.amazon.com, cloudflare.com, akamai.com, digitalocean.com etc \\
\\
Education \& EdTech & 727 &
iitd.ac.in, iitb.ac.in, iitk.ac.in, vit.ac.in, jnu.ac.in, du.ac.in, bits-pilani.ac.in, byjus.com, unacademy.com, physicswallah.live etc \\
\\
E-commerce \& Retail & 351 &
flipkart.com, ajio.com, meesho.com, nykaa.com, bigbasket.com, pepperfry.com, lenskart.com etc \\ \\

Logistics \& Travel & 94 &
delhivery.com, dtdc.in, xpressbees.com, ecomexpress.in, shadowfax.in, porter.in, shiprocket.in, irctc.co.in, makemytrip.com etc \\
\\
FinTech \& Payments & 128 &
razorpay.com, phonepe.com, paytm.com, bharatpe.com, cashfree.cometc \\ \\

Insurance \& Financial Services & 112 & 
hdfclife.com, starhealth.in, policybazaar.com, bajajallianz.com etc \\

Healthcare \& Pharma & 385 &
apollohospitals.com, practo.com, pharmeasy.in, cipla.com, sunpharma.com, drreddys.com, pfizerindia.com etc \\

Media \& Digital Content & 722 &
ndtv.com, thehindu.com, timesofindia.indiatimes.com, news18.com, aajtak.in, hotstar.com, netflix.com etc \\ \\

Social Media Platforms & 42 &
facebook.com, instagram.com, x.com, linkedin.com, reddit.com, discord.com, snapchat.com, threads.com etc \\ \\

Search Engines & 18 &
google.com, bing.com, yahoo.com, duckduckgo.com, yandex.com etc \\\\

Telecom \& Networking & 36 &
airtel.in, jio.com, bsnl.co.in, ericsson.com etc \\\\

Energy \& Critical Infrastructure & 61 & 
ntpc.co.in, iocl.com, adanipower.com, npcil.nic.in etc \\\\

Legal / Advisory / Enterprise Services & 76 &
pwc.in, deloitte.com, kpmg.com, trilegal.com etc \\
\\
Others (Misc) & 27118 &
github.com, stackoverflow.com, medium.com, quora.com etc \\\\
\bottomrule
\end{tabular}
\end{table*}

\subsection{Dataset Collection}

The dataset of analyzed domains was collected using two primary sources:

\begin{itemize}
 \item A total of 31,884 domains were collected from the Tranco domain ranking dataset (\url{https://tranco-list.eu}), which provides a stable and research-oriented list of highly visited Internet domains.

\item A total of 127 domains, including financial institutions, Regional Rural Banks (RRBs), Public Sector Banks, Payment Banks (PBs), Local Area Banks (LABs), Private Sector Banks, Small Finance Banks (SFBs), and foreign banks with a banking presence in India, were collected from the official Reserve Bank of India (RBI) website (\url{https://www.rbi.org.in/scripts/banklinks.aspx}), which lists authorized banking institutions operating in India.
\end{itemize}

The collected domains were merged to form a broader dataset representing multiple layers of critical Internet infrastructure.

\subsection{Sector-wise Distribution of Collected Domains}
A presentation sample of sector wise distribution of collected domains is shown in Table~\ref{tab:sector_domains}. The dataset includes domains from critical sectors such as banking and finance, government services, defense, and energy infrastructure, as well as user-centric platforms including e-commerce, social media, and digital content services. 

In addition, technology-driven sectors such as cloud computing and IT services were incorporated to reflect modern Internet ecosystems. This diverse sectoral distribution enables a comprehensive analysis of security configurations and cryptographic deployments across heterogeneous environments, ranging from enterprise-grade systems to large-scale public platforms. 

The inclusion of both traditional infrastructure and emerging digital services ensures that the dataset captures a realistic and holistic view of current Internet security practices across \textbf{32,011 actively analyzed domains}.

\subsection{Automated TLS Parameter and Certificate Extraction using python}

\begin{algorithm}[htbp]

\caption{Automated Extraction of TLS Versions, Cipher Suites, Key Exchange Mechanisms, HTTP Protocols, Web Servers, and CDN Providers}
\label{alg:tlsframework}

\begin{algorithmic}

\Require List of target domains $X$
\Ensure Structured protocol and cryptographic analysis dataset

\State 1. Normalize all domains in $X$
\State 2. Remove duplicate entries

\ForAll{domain $x \in X$}

    \If{$x$ is invalid}
        \State Skip domain
        \State \textbf{continue}
    \EndIf

    \If{DNS resolution fails}
        \State Mark domain as unreachable
        \State \textbf{continue}
    \EndIf

    \State 3.\textbf{ Establish TLS connection in Chrome browser}

    \State Load target website

    \State Collect browser network security logs

    \State \textit{Extract:}
    \Statex \hspace{1em} Negotiated TLS version
    \Statex \hspace{1em} Cipher suites
    \Statex \hspace{1em} Key exchange mechanism

    \State 5. \textbf{Retrieve and parse X.509 certificate chain}

    \State \textit{Extract:}
    \Statex \hspace{1em} Subject Public key algorithm
    \Statex \hspace{1em} Certificate key size
    \Statex \hspace{1em} Certificate signature algorithm

    \State 4.\textbf{ Perform HTTPS header, CDN, Web server provider analysis from the response header in the network logs }

    \State \textit{Extract }:
    \Statex \hspace{1em} HTTP protocol version
    
    \State Detect CDN provider from response headers
 \Statex \hspace{1em} Web server information

    \Statex \hspace{1em} Alt-Svc headers
   
    \If{Alt-Svc contains HTTP/3 advertisement}
        \State Mark HTTP/3 support as YES
    \Else
        \State Mark HTTP/3 support as NO
    \EndIf

    \If{Post-quantum or hybrid cryptography is detected}
        \State Assign LOW HNDL risk
    \ElsIf{Classical cryptography is detected}
        \State Assign HIGH HNDL risk
    \Else
        \State Assign UNKNOWN HNDL risk
    \EndIf

    \State Aggregate all extracted measurements

    \State Store results in structured CSV dataset

\EndFor

\State \Return Final analysis dataset

\end{algorithmic}
\end{algorithm}

Custom Python scripts (depicted in the Algorithm \ref{alg:tlsframework}) were developed to perform large-scale scanning and data extraction across thousands of domains. The implementation utilized multiple libraries and tools, including:

\begin{itemize}
    \item \textbf{Selenium WebDriver} for browser-based TLS inspection
    \item \textbf{OpenSSL} for handshake-level cryptographic analysis
    \item \textbf{cURL} for HTTP protocol and CDN detection
    \item \textbf{socket, subprocess, threading} for network communication and parallel execution
\end{itemize}

The browser-based component was implemented using \textbf{Selenium WebDriver} with the Chrome DevTools Protocol (CDP). Each domain was accessed using the HTTPS scheme in order to enforce secure communication and initiate a TLS handshake under realistic browser conditions. Chrome performance logs were enabled and collected for every visited domain. From these logs, only \texttt{Network.responseReceived} events were processed. Each response was filtered to ensure that only URLs belonging to the target domain or its subdomains were considered.

For each valid network response, the \texttt{securityDetails} field was extracted. This field provides TLS-specific metadata such as the negotiated TLS protocol version, cipher suite, key exchange method, key exchange group and HTTP protocol. 

To avoid failed scan, before scanning each domain, the browser was reset to a blank page, old performance logs were cleared, Chrome cache was disabled through CDP, and service worker bypass was enabled. This ensured that the collected TLS parameters reflected fresh network communication rather than cached browser responses.

For deeper cryptographic analysis, an OpenSSL-based probing module was implemented. The script used \texttt{openssl s\_client} to establish TLS connections with each target server on port 443 using Server Name Indication (SNI). The handshake output was parsed to extract the negotiated TLS version, cipher suite, key exchange mechanism, and handshake signature algorithm.

Certificate details were extracted using \texttt{openssl x509 } for detailed parsing. The parsed certificate output was used to extract the certificate signature algorithm.

The \textbf{socket} library was used before TLS probing to verify whether a domain resolves successfully. The function \texttt{socket.getaddrinfo()} was used with port 443 and stream socket type. If DNS resolution failed, the domain was skipped. This step prevented unnecessary TLS and HTTP probing against invalid or unreachable domains.

The \textbf{subprocess} library was used to execute external tools such as OpenSSL and cURL from within Python. A wrapper function was implemented around \texttt{subprocess.run()} to capture standard output and standard error, enforce execution timeouts, and return structured status information. Timeout handling was included to prevent the scanner from blocking indefinitely on slow or unresponsive domains. A retry mechanism was also implemented so that timed-out commands could be attempted again before being marked as failed.

The \textbf{cURL} component was used for HTTP-layer and infrastructure analysis. For each domain, cURL was invoked over HTTPS with header-only mode, redirection support, connection timeout, maximum execution time, and custom output formatting. The final response headers were parsed after following redirects. From these headers, the script extracted the \texttt{Alt-Svc} header, server header, HTTP protocol version, and CDN-related indicators.

HTTP/3 support was inferred by checking whether the \texttt{Alt-Svc} header contained an \texttt{h3} advertisement. If such an entry was present, the domain was marked as supporting HTTP/3/QUIC. Otherwise, the HTTP version reported by cURL or the final HTTP status line was mapped to HTTP/1.0, HTTP/1.1, HTTP/2, or HTTP/3. CDN detection was performed using heuristic matching of known response header patterns associated with providers such as Cloudflare, Akamai, Fastly, AWS CloudFront, Google Cloud CDN, Azure CDN, Imperva, and Sucuri.

To support large-scale scanning, the framework used multithreading through Python's \texttt{ThreadPoolExecutor}. Each domain was submitted as an independent scanning task, allowing multiple TLS and HTTP probes to run concurrently. The maximum number of parallel workers was configured as 50. Since multiple threads write results to the same CSV file, a \texttt{threading.Lock()} object was used to ensure thread-safe file writing. This prevented concurrent write conflicts and preserved the integrity of the output dataset.

The complete scanning pipeline follows the sequence: domain normalization, domain validation, DNS resolution, OpenSSL TLS probing, certificate extraction, cURL-based HTTP and CDN analysis, risk assessment, and CSV storage. Each scanned domain produces a structured row containing the domain name, scan status, TLS version, cipher suite, key exchange mechanism, handshake signature algorithm, certificate signature algorithm, subject public key algorithm, certificate key size, HTTP/3 support, \texttt{Alt-Svc} value, web server, and CDN provider.

It is important to note that the extracted TLS parameters represent \textit{negotiated values}. These values are not solely determined by the server. During the TLS handshake, the client advertises supported cipher suites and key exchange groups, and the server selects one mutually compatible configuration. Therefore, the observed cipher suite and key exchange mechanism depend on both client-side configuration, including browser, operating system, and cryptographic libraries, and server-side TLS policy.

By combining Selenium-based browser measurement with OpenSSL and cURL-based active probing, the methodology captures both realistic user-facing TLS behavior and additional cryptographic deployment characteristics required for large-scale security analysis.

\subsubsection{Validation using Wireshark}

To ensure correctness of the extracted data, packet-level validation was performed using Wireshark. TLS handshakes were captured and analyzed to verify:

\begin{itemize}
    \item ClientHello and ServerHello messages
    \item Negotiated cipher suites
    \item Supported cipher suites
    \item Key exchange groups
    \item TLS version agreement
\end{itemize}

This validation step ensured that the values extracted through automated scripts accurately matched the actual network-level communication.

\subsubsection{Verification using Google Chrome Browser Developer Tools}

In addition to automated and packet-level analysis, manual verification was performed using Chrome Developer Tools. The security panel was used to inspect:

\begin{itemize}
    \item TLS version and connection security
    \item Negotiated cipher suites
    \item Certificate details and key sizes
\end{itemize}

This helped confirm that the extracted results were consistent with real-world browser behavior and security indicators.

\section{Results and Analysis}
This section presents the observed deployment patterns and security posture across the analyzed protocols.

\subsection{Quantum Analysis of Protocol Distribution Across Collected Domains}

\begin{table}[h]
\centering
\caption{Percentage Distribution of TLS and QUIC (HTTP/3) Protocols }
\label{tab:protocol_distribution}
\renewcommand{\arraystretch}{1.2}
\begin{tabular}{ccc}
\hline
\textbf{Protocol} & \textbf{Version} & \textbf{Percentage (\%)} \\
\hline
TLS & TLSv1.3 & 52.41\% \\
\hline
TLS & TLSv1.2 & 15.70\% \\
\hline
QUIC & TLSv1.3 (HTTP/3) & 31.89\% \\
\hline
\end{tabular}
\end{table}

The analysis of collected domains reveals that multiple secure communication protocols are actively deployed across the Internet. TLS 1.3 emerges as the most dominant protocol, observed in 16,779 domains, indicating widespread adoption of modern and security-hardened cryptographic standards. In contrast, TLS 1.2 is still present in 5,024 domains, primarily associated with legacy systems and backward compatibility requirements. Additionally, QUIC, implemented via HTTP/3, is observed in 10,208 domains. Since HTTP/3 operates over QUIC and integrates TLS 1.3 directly within its transport layer, all HTTP/3-enabled services were classified under QUIC-based communication in this study.

TLS 1.3 introduces several architectural improvements over earlier protocol versions, including the removal of outdated cryptographic options, simplification of the handshake process, and enforcement of ephemeral key exchange mechanisms, thereby enhancing forward secrecy and resistance against classical network attacks~\cite{rfc8446}. However, despite these advancements, upgrading to TLS 1.3 does not fully address long-term quantum security concerns. Both TLS 1.2 and TLS 1.3 if continue to rely on conventional public-key cryptographic algorithms for certificate authentication and key establishment, then it is computationally vulnerable to quantum algorithms such as Shor’s algorithm, capable of breaking RSA and elliptic curve cryptography~\cite{shor1994}.

\subsubsection{TLS 1.2 vs TLS 1.3 Limitations of TLS 1.2 for QUIC protocol and Post-Quantum Cryptography }

TLS 1.2, defined in RFC 5246, was designed for classical cryptographic environments and supports mechanisms such as RSA and Diffie--Hellman (DH),  for key exchange, along with RSA or ECDSA for authentication~\cite{rfc5246}. In practice, legacy configurations such as RSA-based key exchange are still observed, which lack forward secrecy and are vulnerable to quantum attacks.

TLS 1.3 introduces a simplified and more secure design by enforcing ephemeral key exchange (e.g., X25519, secp256r1), ensuring forward secrecy by default~\cite{rfc8446}. It removes legacy algorithms such as static RSA and CBC-based cipher suites, and supports only authenticated encryption schemes like AES-GCM and ChaCha20-Poly1305. Additionally, TLS 1.3 reduces handshake latency from 2-RTT (in TLS 1.2) to 1-RTT, significantly improving performance.

At the transport layer, both TLS 1.2 and TLS 1.3 are traditionally deployed over TCP, which introduces connection setup delays and suffers from head-of-line (HOL) blocking. QUIC, used in HTTP/3, addresses these limitations by operating over UDP and integrating TLS 1.3 directly into its transport layer~\cite{rfc9000,rfc9001,rfc9114}. This enables faster connection establishment, eliminates HOL blocking through independent streams, and improves both performance and privacy. Due to these architectural requirements, TLS 1.2 is not supported in QUIC.

Despite these improvements, TLS 1.2 remains fundamentally incompatible with post-quantum cryptographic integration. It relies on classical key agreement mechanisms tightly coupled within cipher suites, limiting extensibility. In contrast, post-quantum algorithms such as ML-KEM (Kyber) follow a Key Encapsulation Mechanism (KEM) paradigm and introduce significantly larger key sizes. For example, Kyber-based hybrid key shares (832--1,216 bytes) are substantially larger than classical schemes such as X25519 (32 bytes), leading to potential packet fragmentation and increased transmission overhead.

Furthermore, the rigid handshake structure and higher latency (2-RTT) of TLS 1.2 make it unsuitable for hybrid post-quantum key exchange integration. In contrast, TLS 1.3 introduces a flexible handshake using the \texttt{key\_share} extension, enabling support for hybrid mechanisms such as X25519 combined with ML-KEM~\cite{rfc8446}. As a result, TLS 1.3 provides the necessary architectural foundation for post-quantum cryptography.

Recent deployments have begun adopting hybrid post-quantum key exchange mechanisms to ensure security against both classical and quantum adversaries~\cite{campagna2020,ietfhybrid}. QUIC further facilitates this transition due to its flexible design and faster handshake capabilities, allowing easier integration of evolving cryptographic primitives.

However, a significant portion of Internet infrastructure still relies on legacy protocols such as TLS 1.2, highlighting a gap between protocol evolution and real-world deployment. Achieving full quantum resilience will therefore require a gradual but complete transition toward TLS 1.3 and QUIC-based architectures with native support for post-quantum cryptography.

\subsubsection{Device-wise TLS Configuration Analysis Using Chrome Developer Tools}

To validate protocol deployment one observation was observed at the client level, the security parameters of selected domains was analyzed using Chrome Developer Tools (Security tab) across multiple devices.

\begin{figure}[htbp]
    \centering
    \includegraphics[width=0.5\linewidth]{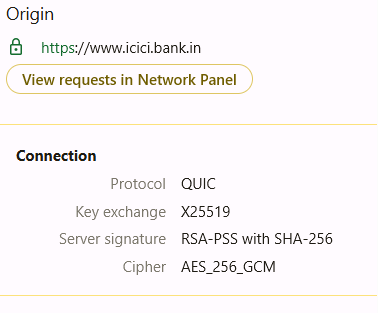}
    \caption{Device 1: QUIC with TLS 1.3 and modern cipher suite observed using Chrome Developer Tools}
    \label{fig:icicidevice1}
\end{figure}

\textbf{Device 1 Observation:}  
Figure \ref{fig:icicidevice1} illustrates the connection to the banking website (icici.bank.in) was established using QUIC, with modern cryptographic parameters including X25519 for key exchange and AES-256-GCM as the cipher. This configuration reflects a secure and optimized connection leveraging QUIC with integrated TLS 1.3~\cite{rfc9001}.

\textbf{Device 2 Observation:}  
On a different device, the same domain (icici.bank.in) was observed to use TLS 1.2, with ECDHE-RSA as the key exchange mechanism and AES-128-GCM as the cipher suite. Despite supporting secure communication, this configuration represents a fallback to an earlier TLS version~\cite{rfc5246}.
      
\begin{figure}[htbp]
    \centering
    \includegraphics[width=0.5\linewidth]{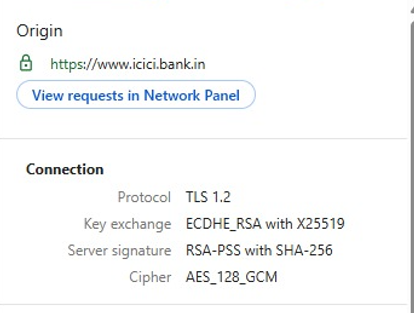}
    \caption{Device 2: TLS 1.2 fallback with legacy-compatible cipher suite observed using Chrome Developer Tools}
    \label{fig:icicidevice2}
\end{figure}

Figure \ref{fig:icicidevice2} shows the variation in TLS versions across devices indicates that protocol negotiation is influenced by client-side factors such as browser version, operating system, and supported cryptographic capabilities. During the handshake process, the server selects the highest mutually supported protocol version between client and server~\cite{rfc8446}.

This demonstrates that even when servers support modern protocols such as QUIC and TLS 1.3, clients may still operate over TLS 1.2 due to compatibility constraints. This highlights the coexistence of modern and legacy protocol deployments in real-world environments.
Although both devices ensure secure communication in practice, TLS 1.3 provides a higher level of security due to enforced modern cryptographic standards, simplified handshake, and resistance to downgrade attacks. Therefore, Device 1 is more secure compared to Device 2.

From a quantum security perspective, this variation introduces a critical concern. Even when servers support advanced protocols, client-side limitations may force fallback to TLS 1.2, which relies on classical cryptographic algorithms. These algorithms, including the RSA and elliptic curve cryptography used in ECDHE, are computationally vulnerable to quantum attacks such as the Shor algorithm~\cite{shor1994}.

This inconsistency implies that achieving quantum resilience is not solely dependent on server-side upgrades, but also requires client-side modernization and widespread adoption of post-quantum cryptographic mechanisms.~\cite{cloudfare,cloudflare_pqc,cloudflare_pqc_support, cloudflare_pqtest} Without such end-to-end upgrades, communication systems remain exposed to long-term threats such as Harvest-Now-Decrypt-Later (HNDL) attacks, where encrypted data intercepted today may be decrypted in the future using quantum computing capabilities~\cite{campagna2020}.

\begin{figure*}[htbp]
    \centering
    \includegraphics[width=0.85\linewidth]{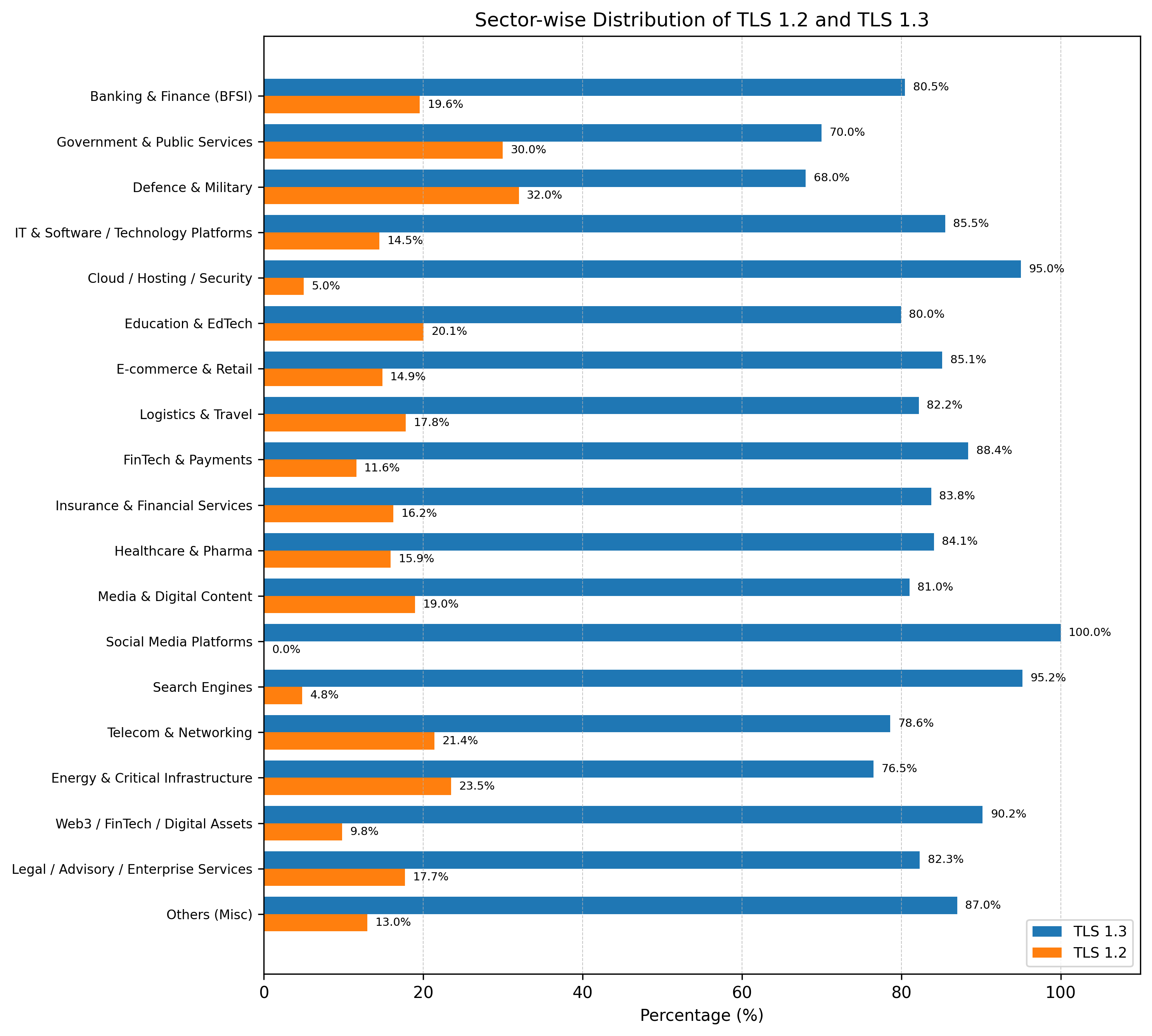}
    \caption{Sector-wise TLS 1.2 and TLS 1.3 Distribution of collected domains}
    \label{fig:placeholder1}
\end{figure*} 
\subsection{Sector-wise TLS Version Distribution and Quantum Vulnerability Analysis}

Figure~\ref{fig:placeholder1} reveals that although TLS 1.3 adoption is dominant across most sectors, a proportion of web servers still operate using TLS 1.2, particularly in critical sectors such as Banking, Government, Defence, Telecom, and Energy.

The continued presence of TLS 1.2 in these sectors can be attributed to practical constraints including legacy infrastructure, compatibility requirements, regulatory dependencies, and the high cost of upgrading production environments. Many financial and government systems prioritize stability and compliance over rapid technological transition, resulting in gradual migration toward TLS 1.3 rather than immediate replacement. NIST also emphasizes that migration to TLS 1.3 provides significant security improvements and helps mitigate vulnerabilities associated with TLS 1.2~\cite{nist_tls13,nist_tls13_visibility,nist_tls13_faq}.

From a security perspective, TLS 1.2 supports a wider range of cryptographic mechanisms, including RSA-based key exchange and older Diffie--Hellman configurations, which increase exposure to weak key exchange and downgrade scenarios~\cite{rfc5246,logjam_nvd,logjam_cisa}. These weaknesses have been exploited in several well-documented attacks such as BEAST, POODLE, Lucky13, and Logjam, which target legacy cipher modes, downgrade mechanisms, and lack of forward secrecy~\cite{rfc7457,poodle2014,lucky13,logjam_nvd}.

TLS 1.3 was specifically designed to eliminate these vulnerabilities. As a result, no widely exploited practical attacks comparable to TLS 1.2 vulnerabilities have been observed against TLS 1.3 in real-world deployments~\cite{rfc8446,nist_tls13_migration}.

Despite these improvements, both TLS 1.2 and TLS 1.3 remain dependent on classical public-key cryptographic algorithms  for authentication, which are vulnerable to quantum attacks. Shor’s algorithm can efficiently break these cryptographic assumptions, enabling future adversaries to decrypt captured encrypted traffic under the Harvest-Now-Decrypt-Later (HNDL) model~\cite{shor1994,nist_pqc}.

The risk is particularly critical in high-value sectors. ENISA reports that public administration accounted for 19\% of observed incidents, while the financial sector represented 8\% in recent threat landscape analyses~\cite{enisa_nis360_2024,enisa_panel_2024}. Additionally, 488 publicly reported incidents were recorded in the financial sector between 2023 and 2024, including multiple large-scale DDoS attacks~\cite{enisa_finance_ddos}. In the public administration sector, data breaches accounted for 17.4\% of incidents, highlighting the impact of confidentiality failures~\cite{enisa_public_admin_2024}.

In practical terms, compromise of TLS-protected communication in these sectors could expose credentials, financial transactions, operational data, and citizen records, leading to fraud, service disruption, identity theft, and loss of trust. The combination of high attack frequency and continued reliance on classical cryptography significantly increases long-term risk.

Overall, while TLS 1.3 improves security and reduces classical attack surfaces, it does not inherently provide quantum resistance. Achieving true quantum resilience requires the integration of post-quantum or hybrid cryptographic mechanisms across both server and client ecosystems. Without such coordinated adoption, critical infrastructure remains exposed to both present-day threats and future quantum-enabled attacks.

\subsection{Quantum Security Analysis of Cipher Suites and key exchanges }

Transport Layer Security establishes secure communication through a handshake process that negotiates cryptographic parameters between the client and the server. During the TLS handshake, the client first sends a \textit{ClientHello} message containing supported protocol versions and cipher suites. The server then responds with a \textit{ServerHello} message selecting a specific cipher suite. The server subsequently provides its digital certificate for authentication, which is typically based on RSA or ECDSA public-key cryptography.\cite{dowling2023tls}

After authentication, a key exchange mechanism such as Elliptic Curve Diffie--Hellman Ephemeral (ECDHE) or Diffie--Hellman Ephemeral (DHE) is used to establish a shared secret between the client and the server. This shared secret is then processed through a key derivation function to generate symmetric session keys. Once these keys are derived, all subsequent communication is encrypted using symmetric encryption algorithms such as AES-GCM or ChaCha20-Poly1305. 

\subsubsection{TLS Cipher Suite Security Classification }

\begin{table}[htbp]
\centering
\caption{Quantum Security Analysis of Observed TLS Cipher Suites}
\label{tab:tls_cipher_quantum}
\scriptsize
\setlength{\tabcolsep}{3pt}
\renewcommand{\arraystretch}{0.5}
\begin{tabular}{lccp{1.5cm}}
\hline\\
\textbf{Cipher Suite} & \textbf{Domains} & \textbf{Enc.} & \textbf{Quantum Impact} \\
\hline
\\
TLS\_AES\_256\_GCM\_SHA384 & 19837 & AES-256 & Grover: 128-bit \\\\\hline\\
TLS\_AES\_128\_GCM\_SHA256 & 6319 & AES-128 & Grover: 64-bit \\
\\
\\\hline\\
TLS\_CHACHA20\_POLY1305\_SHA256 & 127 & ChaCha20 & Grover: 128-bit \\
\\\hline\\

ECDHE-RSA-AES128-GCM-SHA256 & 2305 & AES-128 & Shor break; Grover: 64-bit \\\\\hline\\
ECDHE-RSA-AES256-GCM-SHA384 & 2031 & AES-256 & Shor break; Grover: 128-bit \\\\\hline\\

ECDHE-ECDSA-AES128-GCM-SHA256 & 133 & AES-128 & Shor break; Grover: 64-bit \\\\\hline\\
ECDHE-ECDSA-AES256-GCM-SHA384 & 174 & AES-256 & Shor break; Grover: 128-bit \\\\\hline\\
ECDHE-ECDSA-CHACHA20-POLY1305 & 482 & ChaCha20 & Shor break; Grover: 128-bit \\\\\hline\\

ECDHE-RSA-AES256-SHA & 17 & AES-256 & Shor break \\\\\hline\\
ECDHE-RSA-AES256-SHA384 & 60 & AES-256 & Shor break \\\\\hline\\
ECDHE-RSA-CHACHA20-POLY1305 & 311 & ChaCha20 & Shor break \\\\\hline\\

DHE-RSA-AES128-GCM-SHA256 & 3 & AES-128 & Shor break; Grover: 64-bit \\\\\hline\\
DHE-RSA-AES256-GCM-SHA384 & 12 & AES-256 & Shor break; Grover: 128-bit \\
\\\hline\\
AES\_256\_CBC\_SHA1 & 1 & AES-256 & Weak classical; Grover:128 bit \\
\\
\hline

\end{tabular}
\end{table}
The classification of TLS cipher suites (Table~\ref{tab:tls_cipher_quantum}) highlights a clear distinction between classical and quantum security properties. Most observed cipher suites provide strong classical security through the use of authenticated encryption algorithms such as AES-GCM and ChaCha20, along with forward secrecy mechanisms like ECDHE.
However, despite strong symmetric encryption, all cipher suites rely on classical public-key cryptographic mechanisms for key exchange and authentication. These include RSA and elliptic-curve-based algorithms, which are vulnerable to quantum attacks ~\cite{shor1994}.

As a result, while the majority of cipher suites are classified as strong in the classical security model, they are not secure in the presence of quantum adversaries. Weak classifications within the classical model are primarily associated with legacy constructions such as CBC-mode encryption and SHA-1-based authentication.
This analysis demonstrates that current TLS deployments achieve high levels of security against present-day threats but lack long-term resilience, reinforcing the need for post-quantum cryptographic cipher suites integration.
It is important to note that the observed cipher suites are not inherently insecure. They provide strong security against classical adversaries through authenticated encryption and forward secrecy. However, they are not resistant to quantum attacks due to their reliance on classical public-key cryptographic mechanisms.

\subsubsection{TLS Key Exchange Security Classification }

The study shows that nearly half of the observed key exchange mechanisms (49.3\%) demonstrate quantum-resistant properties through hybrid post-quantum cryptography, while a slightly larger portion (50.7\%) remains vulnerable to quantum attacks.

\begin{table}[htbp]
\centering
\caption{Percentage Distribution of Key Exchange Groups with Quantum Security Impact }
\label{tab:kex_quantum}
\renewcommand{\arraystretch}{1.2}
\begin{tabular}{lcc}
\hline
\textbf{Key Exchange } & \textbf{Percentage (\%)} & \textbf{Quantum Impact} \\
\hline
RSA & 0.09\% & Vulnerable \\\hline\\
prime256v1 (P-256) & 10.33\% & Vulnerable \\\hline\\
secp384r1 & 1.40\% & Vulnerable \\\hline\\
secp521r1 & 0.71\% & Vulnerable \\\hline\\
X25519 & 38.15\% & Vulnerable \\\hline\\
X25519MLKEM768 & 49.30\% & Hybrid (Quantum Resistant) \\\hline\\
X448 & 0.01\% & Vulnerable \\\hline\\

\end{tabular}
\end{table}
This table~\ref{tab:kex_quantum} highlights that the current Internet ecosystem is in a transitional phase towards quantum security. Hybrid key exchange mechanisms offer a robust and practical approach for mitigating emerging quantum threats; however, they cannot be considered absolutely secure or entirely quantum-proof. In cryptographic systems, absolute security is unattainable due to the continuous evolution of attack methodologies and potential implementation weaknesses. Hybrid key exchange integrates classical cryptographic techniques, such as Elliptic Curve Diffie–Hellman (ECDH), with post-quantum cryptographic (PQC) algorithms \cite{nistpqc,ietfhybrid}, such as ML-KEM (Kyber), thereby establishing a dual-layered security framework. This design ensures that even if quantum adversaries compromise the classical component, the PQC component preserves the confidentiality of the communication.

Despite its advantages, hybrid cryptography presents several limitations. Firstly, PQC algorithms are relatively recent and lack the extensive real-world validation that classical algorithms have undergone, leaving room for potential undiscovered cryptanalytic vulnerabilities \cite{nistpqc}. Secondly, the increased complexity of hybrid implementations may introduce software defects and expose systems to side-channel attacks, resulting in unintended information leakage \cite{dowling2023tls}. Thirdly, the “\textit{harvest now, decrypt later}” threat model remains a concern, wherein adversaries may store encrypted data today and decrypt it in the future as more advanced computational capabilities emerge \cite{shor1994}. Additionally, downgrade attacks may coerce systems into reverting to classical-only cryptographic modes, thereby undermining the intended quantum resistance \cite{rfc8446}.

Consequently, while hybrid key exchange mechanisms do not provide absolute quantum security, they represent a significant advancement toward quantum-resilient communication by combining classical and post-quantum cryptographic primitives, ensuring robustness even if one component is compromised \cite{nistpqc,ietfhybrid}.

\noindent

\subsubsection{Packet-Level Validation Using Wireshark}

\begin{figure*}[htbp]
    \centering
    \includegraphics[width=0.65\linewidth]{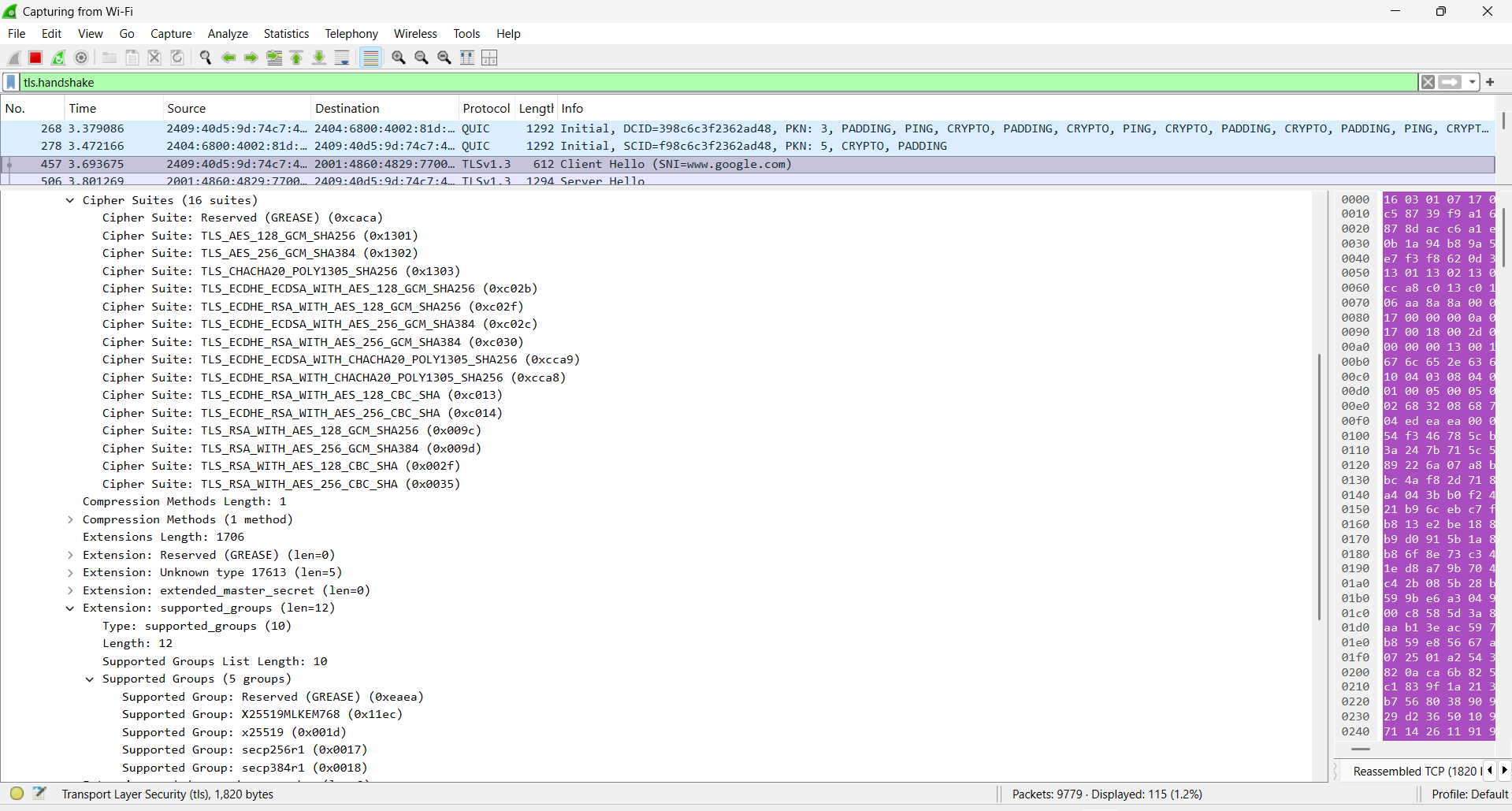}
    \caption{TLS 1.3 Client Hello observed (www.google.com) in Wireshark showing supported cipher suites and supported key exchange groups}
    \label{fig:placeholder}
\end{figure*}

Packet-level inspection of domain: www.google.com using Wireshark reveals the detailed behavior of the TLS 1.3 handshake process. During the Client Hello phase (Fig:~\ref{fig:placeholder}), the client advertises a list of supported cipher suites and cryptographic groups (supported groups extension). These include classical elliptic-curve groups such as X25519 and secp256r1, along with hybrid post-quantum groups such as X25519MLKEM768.

\begin{figure*}[htbp]
    \centering
    \includegraphics[width=0.65\linewidth]{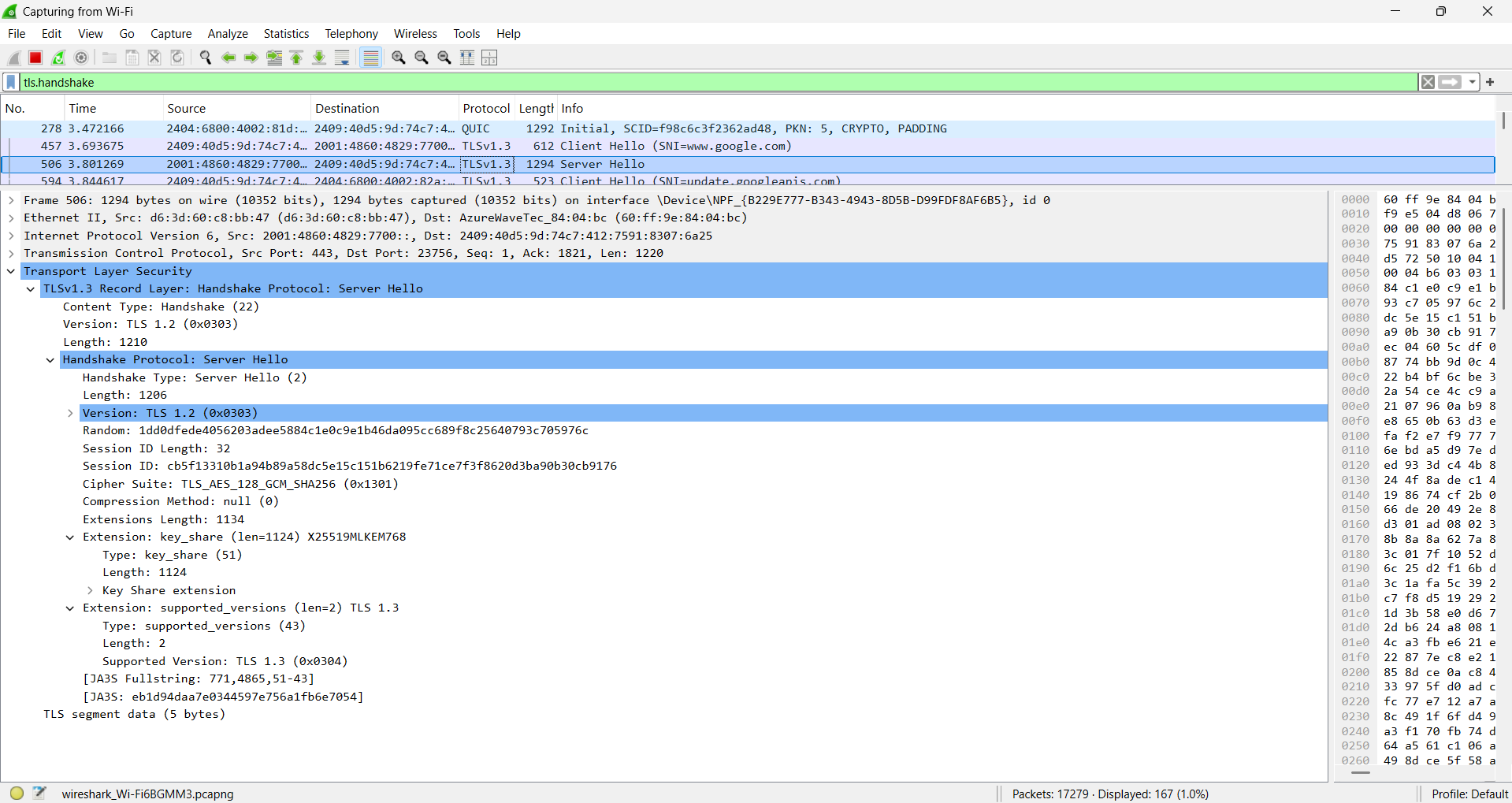}
    \caption{TLS 1.3 Server Hello observed (www.google.com) in Wireshark indicating Negotiated cipher suite TLS\_AES\_128\_GCM\_SHA256 and Negotiated Key Exchange}
    \label{fig:placeholder2}
\end{figure*}

The Server Hello message in Fig.~\ref{fig:placeholder2} indicates the final selection of the cipher suite and key exchange group based on the intersection of client-supported and server-supported configurations. In the observed capture, the negotiated cipher suite is TLS\_AES\_128\_GCM\_SHA256, and the selected key exchange group is X25519MLKEM768.

\begin{figure*}[htbp]
    \centering
    \includegraphics[width=0.65\linewidth]{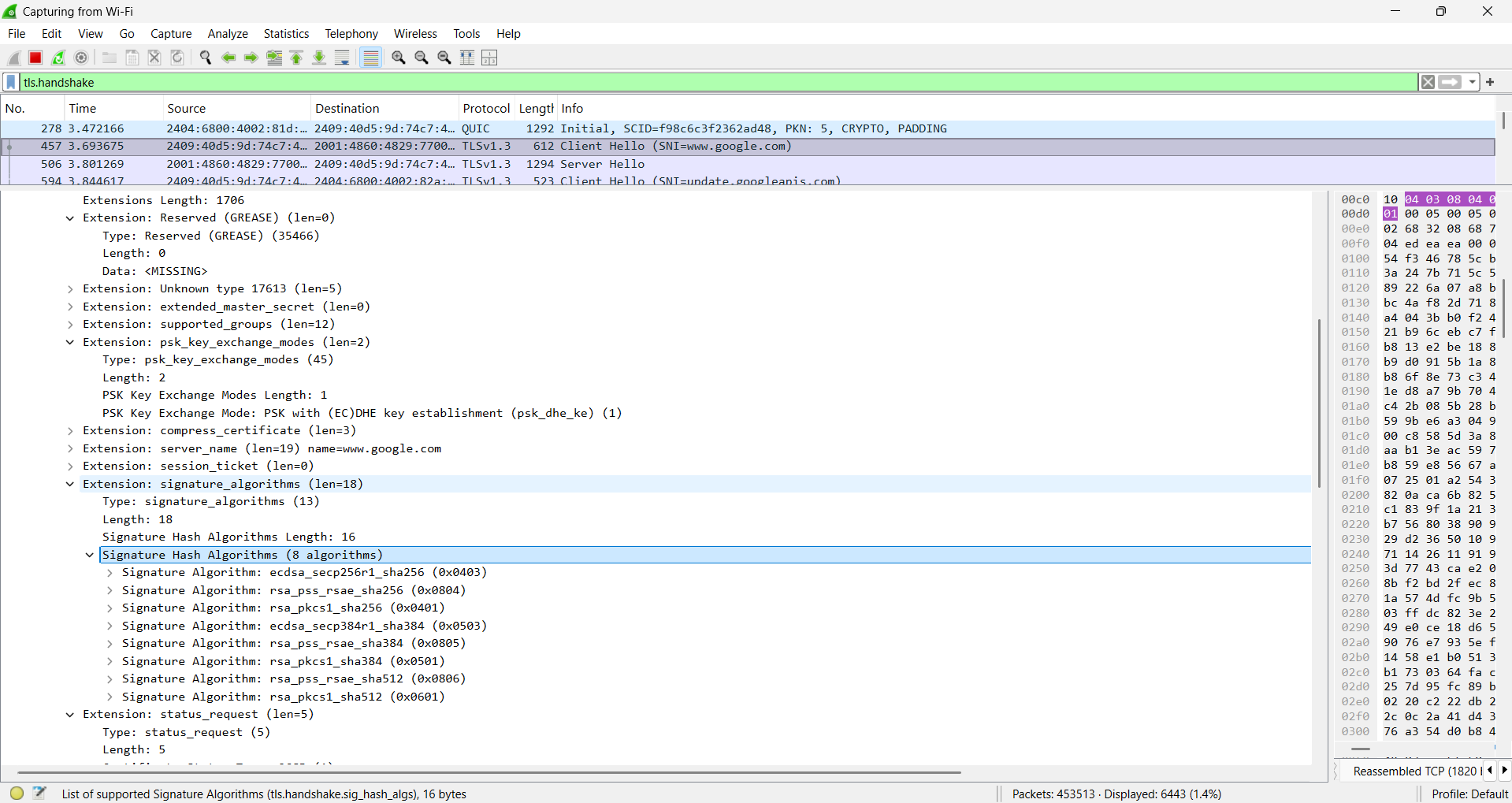}
    \caption{Supported TLS Signature Algorithms  Observed domain: www.google.com in Wireshark Capture}
    \label{fig:googlecert}
\end{figure*}

\subsubsection{Quantum Security Analysis of Certificate Algorithms}

\begin{table}[h]
\centering
\caption{Distribution of Certificate Signature Algorithms with Quantum Security Impact}
\label{tab:signature_quantum}

\scriptsize
\setlength{\tabcolsep}{3pt}
\renewcommand{\arraystretch}{1.0}

\begin{tabular}{lcc}
\hline
\textbf{Signature Algorithm} & \textbf{Percentage} & \textbf{Quantum Security Status} \\
\hline\\
PKCS \#1 SHA-256 with RSA  & 54.05\% & Not Quantum-Safe \\\hline\\

PKCS \#1 SHA-512 with RSA & 2.85\% & Not Quantum-Safe \\\hline\\
X9.62 ECDSA with SHA-256  & 42.32\% & Not Quantum-Safe \\\hline\\
X9.62 ECDSA with SHA-384  & 0.68\% & Not Quantum-Safe \\\hline\\
X9.62 ECDSA with SHA-512 & 0.11\% & Not Quantum-Safe \\

\hline
\end{tabular}
\end{table}
The digital certificates of the target domains were analyzed to extract cryptographic properties, specifically the certificate signature algorithm and the public key size.

The \textbf{signature algorithm} in Table~\ref{tab:signature_quantum} was identified from the X.509 certificate field \textit{signatureAlgorithm}, which specifies the cryptographic algorithm used by the Certificate Authority (CA) to sign the certificate. Commonly observed algorithms include \textit{PKCS \#1 RSA with SHA-256} and \textit{ECDSA with SHA-256/384/512} \cite{rfc5280}. These algorithms reflect the integrity and authenticity mechanism of the certificate.

The \textbf{key size} was observed from the \textit{Subject Public Key Info} field of the certificate, which contains the public key and its associated parameters. For RSA-based certificates, key sizes such as 2048-bit, 3072-bit, and 4096-bit were observed, whereas elliptic curve certificates typically used 256-bit and 384-bit keys \cite{rfc8446}. Therefore, RSA and ECDSA-based certificate algorithms are not quantum-safe due to their reliance on problems vulnerable to Shor’s algorithm \cite{shor1994}. Larger key sizes improve classical security but do not ensure quantum resistance and also no post-quantum algorithms were observed in the dataset \cite{nistpqc}. 

\subsubsection{TLS Handshake Signature Algorithm Analysis using Wireshark}

The TLS handshake was analyzed using Wireshark to observe the supported signature algorithms advertised by the client in the \textit{signature\_algorithms} extension. The captured data includes multiple RSA and ECDSA-based algorithms such as \textit{rsa\_pkcs1\_sha256}, \textit{rsa\_pss\_rsae\_sha256}, and \textit{ecdsa\_secp256r1\_sha256}. These algorithms represent the set of cryptographic options that can be used during certificate verification in the TLS handshake process.

From a quantum security perspective, all observed algorithms are based on RSA and elliptic curve cryptography, which are vulnerable to Shor’s algorithm ~\cite{shor1994}. Therefore, despite being secure in classical environments, these algorithms are classified as not quantum-safe. The absence of post-quantum signature schemes in the TLS handshake highlights the current gap in quantum-resistant deployment in real-world Internet communication.

\subsection{Infrastructure Security Components}

\begin{table}[h]
\centering
\caption{Infrastructure Security Deployment}
\label{tab:infra_security}
\begin{tabular}{lp{4cm}}
\hline
\textbf{Component} & \textbf{Observed Providers} \\
\hline
CDN & Cloudflare, AWS CloudFront, Google Global Edge, Akamai, Fastly \\
Web Server & Apache, Nginx, Cloudflare, Hidden / Obfuscated \\
WAF & Cloudflare WAF, Akamai WAF, Imperva WAF, Sucuri WAF \\
\hline
\end{tabular}
\end{table}
Content delivery networks (CDNs), web application firewalls (WAFs), and managed hosting infrastructures are widely used to improve web security and classical resilience. These technologies help protect services through reverse proxying, traffic filtering, geographic distribution, DDoS mitigation, and application-layer inspection \cite{cloudflare_cdn,akamai_cdn}.

Their presence is an important part of modern Internet defense architecture. However, these components do not directly address the cryptographic challenges introduced by future quantum computing. They primarily enhance availability and operational security, whereas the long-term confidentiality and authenticity of TLS communications depend on the underlying cryptographic algorithms deployed within the protocol itself \cite{rfc8446,shor1994,nistpqc}.

\subsection{CDN, Webserver Infrastructure and Quantum Security Status}

The distribution of CDN providers across the analyzed domains is summarized in Table~\ref{tab:cdn_pqc}.

The analysis of CDN providers ~\ref{tab:cdn_pqc} reveals varying levels of post-quantum cryptographic (PQC) readiness. Major providers such as Cloudflare and AWS CloudFront have already deployed hybrid post-quantum key exchange mechanisms (x25519 with ML-KEM/Kyber), indicating strong readiness for future quantum threats~\cite{cloudflare_pqc,aws_pqc}. however, Cloudflare implements hybrid post-quantum key exchange mechanisms in default mode to strengthen TLS security against future quantum threats.
however, hybrid mechanism X25519+ML-KEM-768 is widely adopted in default mode across major browsers, including Chrome 131+, Firefox 132+ (desktop) and 145+ (Android), Safari 26+, Edge 131+, as well as Opera, Brave, and Tor Browser 15.0+. It is also supported system-wide in modern Apple operating systems such as iOS 26 and macOS Tahoe 26. In addition to browsers, this hybrid approach is integrated into modern cryptographic libraries and tools, including Go 1.24+, OpenSSL 3.5.0+, Node.js (v24.5.0+ and v22.20.0+), BoringSSL, GnuTLS (with leancrypto or liboqs), rustls 0.23.22+, and Botan 3.7.0+. It is also supported in server and networking software such as NGINX (compiled with OpenSSL 3.5+), Caddy 2.10.0+, Traefik, rpxy 0.9.4+, and implementations from Open Quantum Safe (liboqs and oqs-provider) ~\cite{cloudflare_pqc_support}. Additionally, support is available in Zig 0.14.0+ and Cloudflare's Go fork. Google Cloud CDN also demonstrates hybrid PQC adoption in controlled or internal key exchange environments~\cite{google_pqc}.

\begin{table}[h]
\centering
\caption{CDN Providers with PQC readiness Status}
\label{tab:cdn_pqc}

\scriptsize
\setlength{\tabcolsep}{3pt}
\renewcommand{\arraystretch}{1.0}
\begin{tabular}{lcc}
\toprule
\textbf{Provider} & \textbf{Percentage (\%)} & \textbf{Key Exchange } \\
\midrule

Cloudflare & 37.97 & ML-KEM Hybrid (Kyber) \\
AWS CloudFront & 7.71 & ML-KEM \& ML-DSA \\
Google Cloud CDN / GFE & 1.76 & ML-KEM \& ML-DSA \\
Akamai & 2.39 & ML-KEM \\
Fastly & 3.17 & ML-KEM(Kyber) in TLS 1.3 \\
Azure CDN & 0.79 & Depends on Partner CDN \\
Imperva & 0.67 & Hybrid PQC (X25519MLKEM768) \\
Sucuri & 0.16 & (No Direct PQC Yet found,TLS 1.3) \\
Hidden / Not Detected & 45.37 & Unknown \\

\bottomrule
\end{tabular}
\end{table}

In contrast, providers such as Akamai have also introduced hybrid post-quantum key exchange mechanisms (e.g., X25519MLKEM768) within TLS~1.3 at the edge, enabling optional PQC support for client connections~\cite{akamai_pqc}. This reflects an active transition phase, where NIST-standardized PQC mechanisms are being gradually integrated into production environments, although full-scale deployment is still ongoing~\cite{akamai_pqc,nist_pqc,nistpqc}.

Similarly, Fastly has begun integrating x25519ML-KEM768 into TLS~1.3 handshakes, indicating a transition toward quantum-resistant infrastructure~\cite{fastly_pqc}. Imperva also demonstrates partial PQC readiness through hybrid TLS key exchange (X25519+ML-KEM-768), ensuring backward compatibility while enabling quantum-resistant connections when supported by clients~\cite{imperva_pqc}.
Sucuri actively monitors developments in post-quantum cryptography (PQC) and relies on modern, industry-standard TLS configurations and infrastructure practices through its security stack and upstream providers. While large-scale migration to PQC is still evolving across the industry, Sucuri emphasizes protecting customer traffic and maintaining alignment with current cryptographic best practices as part of its long-term security roadmap.~\cite{sucuri}.

A significant portion of domains rely on hidden or unidentified CDN configurations, making their quantum readiness difficult to assess. Overall, these observations highlight that while leading CDN providers are actively progressing toward quantum-safe infrastructure, widespread adoption across the Internet remains incomplete and uneven~\cite{nist_pqc,nistpqc}.

\begin{table*}[htbp]
\centering
\caption{Percentage Distribution of Observed Web Server Infrastructure with Quantum Readiness}
\label{tab:webserver_percentage}
\renewcommand{\arraystretch}{1.1}

\resizebox{\linewidth}{!}{
\begin{tabular}{lp{6cm}cc}
\hline
\textbf{Category} & \textbf{Technology} & \textbf{Percentage (\%)} & \textbf{PQC Readiness} \\
\hline

Web Server & Apache (all variants) & 16.25\% & Classical ;(Dependent on underlying TLS libraries such as OpenSSL; PQC possible via hybrid extensions) \\

Web Server & Nginx (all variants) & 19.37\% & Classical; ( Hybrid PQC supported when compiled with OpenSSL $\geq$ 3.5 or OQS provider) \\

Web Server & Microsoft IIS & 1.09\% & Classical; (Hybrid PQC support$\geq$ Windows server 2025 , Windows 11 ; and future updates) \\

Web Server & LiteSpeed, Caddy, Kestrel, Gunicorn & 1.16\% &  (Hybrid PQC support experimental; e.g., Caddy via Go TLS hybrid implementations) \\

\hline
CDN / Edge & Cloudflare & 35.93\% & (default X25519MLKEM768 in TLS 1.3) \\

CDN / Edge & CloudFront, Akamai, GFE, Amazon S3 & 4.69\% &  (default X25519MLKEM768 in TLS 1.3 ) \\

\hline
Proxy / Load Balancer & ELB, Envoy, HAProxy & 1.59\% & Indirect (Hybrid PQC support depends on TLS libraries ) \\

\hline
Hidden / Unknown & Masked or Not Identified & 35.56\% & Unknown (Insufficient visibility to assess PQC readiness) \\

\hline
\end{tabular}
}

\end{table*}

Table~\ref{tab:webserver_percentage} shows that a significant portion of the observed infrastructure is either hidden or served through CDN and proxy layers, indicating that modern web architectures increasingly obscure origin server details for security and performance optimization. 

NGINX does not natively implement post-quantum cryptography; instead, hybrid PQC support is enabled through underlying cryptographic libraries such as OpenSSL (version $\geq$ 3.5) or the Open Quantum Safe (OQS) provider, highlighting that PQC adoption in web servers is largely dependent on external TLS implementations rather than the server software itself~\cite{nginx_pqc}.
Microsoft has introduced post-quantum cryptographic (PQC) support in Windows Server 2025, Windows 11, and .NET 10 through the integration of NIST-standardized algorithms such as ML-KEM and ML-DSA. These algorithms enable quantum-resistant key exchange and digital signatures, respectively, and are designed to mitigate emerging threats such as "harvest now, decrypt later." However, current deployment follows a hybrid approach, combining classical and post-quantum mechanisms to ensure backward compatibility and gradual ecosystem transition~\cite{microsoft_pqc}.

\subsection{Sector-wise Quantum Risk Heatmap}
The quantum risk heatmap~\ref{tab:quantum_heatmap} highlights significant disparities in post-quantum readiness across different sectors.

\begin{table}[htbp]
\centering
\caption{Sector-wise Quantum Risk Heatmap}
\label{tab:quantum_heatmap}
\renewcommand{\arraystretch}{1.1}
\setlength{\tabcolsep}{6pt}
\scriptsize
\begin{tabular}{|l|c|c|c|}
\hline
\textbf{Sector}  & \textbf{PQC Adoption} & \textbf{Quantum Risk} \\
\hline

BFSI & Low & \cellcolor{red!30} High \\
Government & None & \cellcolor{red!40} High \\
Defence  & None & \cellcolor{red!60} Very High \\
IT/Tech  & Partial & \cellcolor{orange!40} Medium \\
Cloud/CDN & High & \cellcolor{green!40} Low \\
Social Media & High & \cellcolor{green!40} Low \\
E-commerce & Medium & \cellcolor{orange!30} Medium \\
Enterprise (Microsoft)  & None & \cellcolor{red!30} High \\
Security Vendors  & None & \cellcolor{red!30} High \\
Media  & High & \cellcolor{green!40} Low \\
Search Engines  & High & \cellcolor{green!50} Low \\
Telecom  & None & \cellcolor{red!30} High \\
Energy & None & \cellcolor{red!30} High \\
Legal/Advisory & None & \cellcolor{red!30} High \\

\hline
\end{tabular}
\end{table}

Technology-driven sectors such as Cloud/CDN, Social Media, and Search Engines demonstrate low quantum risk due to the adoption of hybrid post-quantum key exchange mechanisms (e.g., ML-KEM combined with X25519). These sectors are leading the transition toward quantum-resilient communication.

In contrast, sectors such as BFSI, Government, Defence, and Telecom exhibit high quantum risk due to continued reliance on classical public-key cryptographic algorithms such as RSA and ECDSA. These algorithms are vulnerable to future quantum attacks, particularly under the harvest-now-decrypt-later (HNDL) threat model. Enterprise and security-related services show a mixed trend, with modern transport protocols but limited adoption of post-quantum cryptography, indicating transitional deployment phases.

Overall, the heatmap reveals that quantum readiness is not uniform and is strongly influenced by sector-specific priorities, with performance-driven platforms leading adoption and critical infrastructure lagging behind.

\section{Discussions and Conclusions}
One of the key quantitative findings of this study is that 50.7\% of the analyzed domains remain fully vulnerable to quantum attacks due to their continued reliance on classical cryptographic mechanisms, while 49.3\% demonstrate partial post-quantum readiness through the deployment of hybrid key exchange mechanisms. These results indicate that the Internet ecosystem is currently in a transitional phase of post-quantum cryptographic (PQC) adoption. A major driver of this transition is the growing deployment of ML-KEM-768 (Kyber-768), standardized by NIST under FIPS~203. Its favorable balance between security and performance, combined with integration into modern TLS implementations, browsers, cloud platforms, and content delivery networks, has enabled practical Internet-scale deployment. Previous studies and large-scale deployments by providers such as Google and Cloudflare \cite{cloudflare_pqc_support, google_pqc_standards} have further demonstrated that hybrid key exchange schemes based on X25519 and ML-KEM-768 introduce minimal performance overhead while providing protection against both classical and quantum adversaries.

Despite the availability of standardized PQC algorithms, including ML-KEM (FIPS~203), ML-DSA (FIPS~204), and SLH-DSA (FIPS~205), large-scale deployment remains incomplete. The primary barriers are no longer the absence of suitable cryptographic standards but the operational and compatibility challenges associated with migration. Many Internet-facing systems continue to rely on legacy hardware, software, and security appliances that require extensive testing, validation, and interoperability assessment before cryptographic upgrades can be deployed. In addition, PQC adoption is closely tied to support within cryptographic libraries, operating systems, browsers, and application frameworks, causing deployment to progress at the pace of broader software ecosystem updates.

Sector-wise observations reveal that adoption rates vary according to organizational requirements and infrastructure maturity. Highly regulated sectors, including banking, financial services, government, defence, and telecommunications, generally exhibit slower migration due to strict compliance requirements, certification procedures, and long infrastructure life cycles. In contrast, cloud-native organizations and large technology providers have adopted hybrid PQC mechanisms more rapidly through upgrades within cloud and CDN infrastructures.

The continued prevalence of TLS~1.2 further highlights the challenges associated with cryptographic modernization. Although TLS~1.3 provides a more suitable framework for integrating hybrid post-quantum key exchange mechanisms, many organizations continue to support TLS~1.2 to maintain compatibility with legacy systems, embedded devices, older operating systems, and long-lived enterprise applications. Consequently, a substantial portion of Internet communications remains dependent on classical cryptographic mechanisms and cannot fully benefit from current hybrid PQC deployments.
Collectively, these findings suggest that the primary barriers to PQC adoption are no longer the absence of standardized algorithms but rather the operational, economic, and compatibility challenges associated with large-scale migration.

However, a critical limitation identified in this study is that certificate-based authentication remains entirely classical. TLS certificates continue to rely on RSA (56.9\%) and ECDSA (43.1\%) signature schemes, and no instances of hybrid or post-quantum digital signature algorithms were observed in X.509 certificates across the dataset. This finding reflects the current state of the public key infrastructure (PKI) ecosystem, where support for post-quantum certificate issuance, certificate transparency logging, certificate authority workflows, and browser trust stores remains under active development.

This creates a fundamental security gap. While hybrid key exchange mechanisms improve confidentiality against future quantum adversaries, authentication remains dependent on quantum-vulnerable signature schemes. Under a quantum threat model, an adversary capable of breaking classical signature algorithms could forge certificates, impersonate legitimate servers, and conduct man-in-the-middle attacks. Consequently, hybrid key exchange alone is insufficient to provide end-to-end quantum security.

At the infrastructure level, a clear divide exists between early adopters and traditional systems. This trend is currently concentrated among major cloud, CDN, and edge-service providers, whereas conventional web infrastructures continue to rely predominantly on classical TLS deployments. In most cases, PQC support is not implemented directly at the application layer but is inherited from underlying cryptographic libraries such as OpenSSL or BoringSSL, which significantly influences the pace of adoption.

Another important observation is the dynamic nature of TLS configuration negotiation. Cryptographic parameters are selected during the handshake based on both client-side and server-side capabilities. As a result, even when servers support hybrid PQC, clients without such support may fall back to classical cryptographic algorithms.

This fallback behavior is further validated through device-level observations and sector-wise analysis. For example, one device successfully established a QUIC connection using TLS~1.3 with modern cryptographic primitives, while another device connected to the same server using TLS~1.2 with classical cipher suites. Sector-wise data further shows that critical domains such as Government, Defence, and Telecom exhibit higher reliance on TLS~1.2 compared to cloud-centric sectors.

From a quantum security perspective, this fallback has significant implications. TLS~1.2 configurations predominantly rely on classical key exchange mechanisms such as ECDHE-RSA, which are vulnerable to Shor's algorithm, whereas TLS~1.3 enables hybrid PQC integration. Consequently, fallback scenarios can negate quantum-resistant benefits and lead to inconsistent security guarantees across connections.

If these implications are not addressed, the Internet's trust model faces substantial systemic risk under a quantum threat. First, the continued use of RSA and ECDSA in certificates, combined with the absence of PQC-based certificate deployment, may enable \textit{certificate forgery} once large-scale quantum computing becomes practical. Such capabilities would allow adversaries to impersonate legitimate services and perform undetectable man-in-the-middle attacks. Second, the prevalence of classical key exchange mechanisms and TLS~1.2 fallback paths facilitates \textit{retrospective decryption} under the Harvest-Now-Decrypt-Later (HNDL) threat model, exposing previously captured sensitive data, including financial transactions, authentication credentials, and government records. Third, inconsistent client–server negotiation leads to \textit{downgrade-driven security fragmentation}, where even PQC-capable servers fall back to classical sessions for legacy clients, nullifying quantum-resistant benefits in practice. Finally, critical sectors such as BFSI, government, and defence accumulate concentrated risk due to slower upgrade cycles, increasing the likelihood of large-scale data breaches, identity fraud, and loss of service integrity. Without coordinated migration of both key exchange mechanisms and certificate-based authentication to post-quantum primitives, these risks will persist even as partial PQC adoption continues to grow.

Overall, the coexistence of classical and hybrid cryptographic mechanisms reflects a gradual migration rather than a complete transition. Although the foundation for PQC adoption has been established through hybrid key exchange deployment, large-scale implementation remains incomplete, and systems continue to face long-term risks associated with Harvest-Now-Decrypt-Later (HNDL) attacks.

In conclusion, hybrid key exchange mechanisms represent a significant step toward partial quantum resilience but do not provide complete security due to the continued reliance on classical certificate-based authentication. Achieving full end-to-end quantum security requires the migration of both key exchange mechanisms and certificate infrastructures to standardized post-quantum cryptographic algorithms.

Future work should focus on the integration of PQC-based certificates, evaluation of fully post-quantum TLS handshakes, performance and interoperability assessment of post-quantum authentication mechanisms, and large-scale deployment strategies. Without such coordinated efforts, critical Internet infrastructure will remain vulnerable to quantum-enabled adversaries.

\bibliography{references}

\end{document}